\documentclass[final]{aipproc}
\usepackage{amssymb}
\usepackage{amsmath}

\layoutstyle{6x9}

\begin{document}

\title{Multiquark Correlations in Light Mesons and Baryons from Holographic
QCD}


\author{Hilmar Forkel}{address={Institut für Physik, 
Humboldt-Universität zu Berlin, D-12489 Berlin, Germany}}

\begin{abstract}
A hadron's multiquark content reflects itself in the quark composition of
the interpolator with which it has maximal overlap. The AdS/CFT dictionary
translates the anomalous dimension of this interpolator into a mass
correction for the corresponding dual mode. Hence such bulk-mass corrections
can carry holographic information on multiquark correlations. Two prominent
examples are studied by implementing this robust and universal 
mechanism into AdS/QCD gravity duals. In the baryon sector bulk-mass
corrections are used to describe systematic good (i.e. maximally attractive)
diquark effects. The baryon sizes are predicted to decrease with increasing
good-diquark content, and the masses of all 48 observed light-quark baryon
states are reproduced with unprecedented accuracy. Our approach further
provides the first holographic description of a dominant tetraquark
component in the lowest-lying scalar mesons. The tetraquark ground state
emerges naturally as the lightest scalar nonet whereas higher excitations
become heavier than their quark--antiquark counterparts and are thus likely
to dissolve into the multiparticle continuum.
\end{abstract}

\maketitle

\section{Introduction}

The main objective of the AdS/QCD program \cite{adsqcd,revs2} is to describe
strong-interaction and especially vacuum and hadron physics holographically
by means of a gravitational dual dynamics in $5D$ spacetime (the
\textquotedblleft bulk\textquotedblright ). The latter has an IR\ deformed
AdS$_{5}$ geometry 
\begin{equation}
ds^{2}=g_{MN}dx^{M}dx^{N}=e^{2A\left( z\right) }\frac{R^{2}}{z^{2}}\left(
\eta _{\mu \nu }dx^{\mu }dx^{\nu }-dz^{2}\right)  \label{met}
\end{equation}%
($R$ is the AdS$_{5}$ curvature radius, $z$ parametrizes the fifth
dimension, $A\left( z\right) \overset{z\rightarrow 0}{\longrightarrow }0$)
and generally contains other fields (e.g. dilatons) as well. Essential
long-term goals of this \textquotedblleft bottom-up\textquotedblright\
approach are to construct an at least reasonably close approximation to the
QCD dual and to supply top-down string theory approaches with specific ideas
for the relevant dynamics. Over the last years, this program has made
significant progress \cite{AdSQCDNewer}.

In the present talk we discuss multiquark correlations in mesons and baryons
from a holographic perspective. In particular, we review a systematic and
universal extension of the AdS/QCD framework which describes such
correlations holographically. To motivate our approach, we recall that the
bulk mode $\varphi $ dual to a given hadron is associated with the
gauge-theory interpolater with which it has maximal overlap. The scaling (or
twist) dimension $\Delta $ of this interpolator sets the UV boundary
condition \cite{revs1} 
\begin{equation}
\varphi \left( x;z\right) \overset{z\rightarrow 0}{\longrightarrow }\varphi
^{\left( 0\right) }\left( x\right) z^{f\left( \Delta \left( m_{5}R\right)
\right) }  \label{bc}
\end{equation}%
where $f$ is a known, hadron-dependent function. The condition (\ref{bc}) is
imposed on the (at small $z$ leading) solutions of the\ bulk field equations
by adjusting the mass term. As a result, the scaling dimension $\Delta
\left( m_{5}R\right) $ becomes a function of the $5D$ bulk mass $m_{5}$.

In general, there exist several gauge invariant interpolators $\eta _{i}$
with equal quantum numbers and the same classical dimension 
$\Delta _{\text{cl}}$. Those combine the fundamental fields in different 
ways but can couple to the same hadron state $\left\vert h\right\rangle $. 
Hence the overlaps $%
\left\langle 0\left\vert \eta _{i}\right\vert h\right\rangle $ contain
subtle infomation on the hadron's structure. In particular, the field
composition of the interpolator with maximal overlap is expected to most
closely represent the coupling among the valence partons of the hadron. In
our context it is crucial to note that such structural differences manifest
themselves also in different \emph{anomalous} dimensions\footnote{%
This is borne out in perturbative-QCD calculations of $\gamma $, e.g. for
diquark operators \cite{kle11}.} $\gamma _{i}\left( \mu \right) $ of the $%
\eta _{i}$. The boundary condition (\ref{bc}) then translates the $\gamma
\left( \mu \right) $ of a given interpolator $\eta $ into a bulk mass
correction $\Delta m_{5}\left( z\right) $ for the corresponding dual modes.
(The $z$ dependence is inherited from the scale dependence of $\gamma \left(
\mu \right) $ since $\mu \sim 1/z$.) Hence such corrections can encode
multiquark correlation effects \cite{for10,for09} inside hadrons\footnote{%
To a reasonable approximation those are probably negligible in quite a few
hadrons, but not in all \cite{for10}. Anomalous-dimension induced
corrections in a different context were explored in Ref. \cite{veg10}.}.

The practical implementation of the above corrections into bottom-up duals
is not without challenges, however. For example, it is not \emph{a priori}
obvious how to assign unique hadron states to the interpolators. QCD
information on anomalous dimensions of hadronic interpolators in the
infrared is still scarce, furthermore. A particular challenge is the
necessarily naive AdS/QCD extrapolation of the multiquark physics with its
pronounced $N_{c}$ dependence from large $N_{c}$ to $N_{c}=3$ (cf. Ref. \cite%
{for10}). Major benefits, on the other hand, are that the outlined mechanism
will work in any of the current AdS/QCD duals, both non-dynamical and 
dynamical (i.e. backreacted), and that the results will be rather independent
of the chosen dual. In the following we invoke this mechanism to describe
diquark effects in the light-quark baryon spectrum \cite{for09} and
tetraquark correlations inside the lightest scalar mesons \cite{for10}.

\section{Diquark correlations in light-quark baryons}

The arguably most prominent pattern in measured hadron spectra consists of
(approximately) linear trajectories with universal slopes on which the
square masses $M^{2}$ of excited states organize themselves as a function of
both angular momentum $L$ (or alternatively total spin $J$) and radial
excitation level $n$. A rather minimal \textquotedblleft metric
soft-wall\textquotedblright\ (ms) gravity dual \cite{for07} generates such
trajectories in the light-quark meson and baryon spectra. In the baryon
sector, in particular, it predicts 
\begin{equation}
M_{n,L}^{\left( \mathrm{ms}\right) 2}=4\lambda ^{2}\left( n+L+\frac{3}{2}%
\right) .  \label{mms}
\end{equation}%
While Eq. (\ref{mms}) describes all experimental data for the delta
resonance masses within errors \cite{kle08}, systematic deviations remain
noticeable in the nucleon sector. In Ref. \cite{for09} the latter were
related to the fraction $\kappa $ of \textquotedblleft
good\textquotedblright\ (i.e. maximally attractive) diquarks in the baryons'
quark-model wave function (for recent related work see e.g. \cite{diqbar}).
These diquark correlations are expected to generate the mass correction%
\begin{equation}
\Delta M_{\kappa }^{2}=-2\left( M_{\Delta }^{2}-M_{N}^{2}\right) \kappa 
\label{dcmf}
\end{equation}%
which solely depends on the resonances' diquark content ($\kappa =0$ for
deltas). In Ref. \cite{for09} we have analyzed whether the
anomalous-dimension-based implementation of multiquark effects is able to
generate such corrections in the metric soft-wall dual. We start from the
two independent leading-twist nucleon interpolators 
\begin{equation}
\eta _{\mathrm{pd}}=\varepsilon _{abc}\left( u_{a}^{T}Cd_{b}\right) \gamma
^{5}u_{c},\text{ \ \ \ \ \ }\eta _{\mathrm{sd}}=\varepsilon _{abc}\left(
u_{a}^{T}C\gamma ^{5}d_{b}\right) u_{c}  \label{ni}
\end{equation}%
of QCD with $\Delta =9/2$ which contain in $\eta _{\mathrm{pd}}$ a
pseudoscalar and in $\eta _{\mathrm{sd}}$ a \textquotedblleft
good\textquotedblright\ scalar diquark (and for $L>0$ additional covariant
derivatives). The interpolators (\ref{ni}) are expected to have enhanced
overlap with nucleon states of equivalent diquark content. Hence we
introduce three bulk spinor fields $\Psi ^{\left( \kappa \right) }$ dual to
the linear combinations of the interpolators (\ref{ni}) with $\kappa =0,$ $%
1/4$ and $1/2,$ and thus to nucleons with the same diquark substructure.
Following our above reasoning, the $\Psi ^{\left( \kappa \right) }$ are
solutions of the bulk Dirac equation 
\begin{equation}
\left[ ie_{A}^{M}\Gamma ^{A}\left( \partial _{M}+2\partial _{M}A^{\left( 
\text{ms}\right) }\right) -m_{5}^{\left( \kappa \right) }\right] \Psi \left(
x,z\right) =0  \label{de}
\end{equation}%
(where $\Gamma ^{A}$ are $5D$ Dirac matrices and $e_{M}^{A}=\delta
_{M}^{A}\exp A^{\left( \text{ms}\right) }$ f\"{u}nfbeins) in the geometry (%
\ref{met}) with the warp factor $A^{\left( \text{ms}\right) }\left( z\right)
=\ln \left[ \left( R/z\right) \left( 1+\lambda ^{2}z^{2}/m_{5}^{\left( 
\text{ms}\right) }/R\right) \right] $ \cite{for07}. The bulk masses%
\begin{equation}
m_{5}^{\left( \kappa \right) }=m_{5}^{\left( \text{ms}\right) }+\Delta
m_{5}^{\left( \kappa \right) }=\frac{L+\Delta m_{5}^{\left( \kappa \right)
}R+1}{R},\text{ \ \ \ \ }\kappa \in \left\{ 0,1/4,1/2\right\}   \label{m5}
\end{equation}%
originate from $m_{5}^{\left( \text{ms}\right) }=\left( L+1\right) /R$ \cite%
{for07}, which accounts for the \emph{classical} twist dimension $\bar{\tau}%
=L+3$ of the baryon interpolators (\ref{ni}), and from their anomalous
dimensions $\gamma _{\kappa }=\Delta m_{5}^{\left( \kappa \right) }$.
Equation (\ref{m5}) ensures that the chirally-odd components of $\Psi
^{\left( \kappa \right) }$ satisfy the AdS/CFT boundary condition (\ref{bc}%
). Although no QCD information on the nonperturbative $\gamma \left( \mu
\right) $\ is currently available in the IR (for $\mu <2$ GeV \cite{goe09})
one can check whether the phenomenologically successful modification (\ref%
{dcmf}) may be obtained from a judiciously chosen bulk-mass correction $%
\Delta m_{5}^{\left( \kappa \right) }$. This indeed turns out to be the case 
\cite{for09}. In fact, with%
\begin{equation}
\Delta m_{5}^{\left( \kappa \right) }=\frac{\Delta M_{\kappa }^{2}}{4\lambda
^{2}R}  \label{dm5}
\end{equation}%
the normalizable solutions of Eq. (\ref{de}) generate the desired eigenvalue
spectrum%
\begin{equation}
M_{n,L}^{2}=4\lambda ^{2}\left( n+L+\frac{3}{2}\right) -2\left( M_{\Delta
}^{2}-M_{N}^{2}\right) \kappa .  \label{dcbms}
\end{equation}%
(Resolution-dependent bulk-mass corrections which reproduce Eq. (\ref{dcbms}%
) can be constructed as well \cite{for09}.) Equation (\ref{dcbms}) is in
excellent agreement with all existing data. The eigenmode solutions can be
obtained analytically and reveal that modes corresponding to larger $\kappa $
(with identical $n$, $L$) extend less into the fifth dimension. This
translates into a smaller size of baryons with increased attraction from the
good-diquark channel \cite{for09}.

\section{Light scalar mesons as tetraquarks}

The lightest scalar meson nonet \cite{clo02} has long been suggested to
contain a dominant tetraquark component \cite{jaf77}. This at present
arguably favored interpretation \cite{mai07} requires an exceptionally
strong four-quark binding. The underlying dynamics, able to push the
tetraquark ground-state mass below the mass of the lightest scalar $\bar{q}q$
state, must similarly be of exceptional origin. This is reflected in the
holographic description of spin-0 mesons. Indeed, a straightforward
extension of the conventional approach, based on quark--antiquark
interpolators \cite{veg08}, to $\bar{q}^{2}q^{2}$ interpolators\ would
result in four-quark states which are heavier, not lighter than the $\bar{q}%
q $ ground state. This is because the mass $m_{5}$ of the scalar bulk-mode
solutions satisfies \cite{revs1}

\begin{equation}
m_{5}^{2}R^{2}=\Delta \left( \Delta -4\right)  \label{m}
\end{equation}%
to meet the boundary condition (\ref{bc}). Hence the larger classical
dimension $\Delta $ of interpolators with a larger quark-field content
generates larger bulk-mode and meson masses.

To put these and the following considerations into an explicit dynamical
context, we adopt the popular dilaton-soft-wall dual of Ref. \cite{kar06}
(which is based on the AdS$_{5}$ metric (\ref{met}) with $A\equiv 0$ and a
quadratic dilaton\ $\Phi \left( z\right) =\lambda ^{2}z^{2}$). The bulk
modes dual to scalar mesons in this background solve the radial
Klein-Gordon equation 
\begin{equation}
\left[ -\partial _{z}^{2}+V\left( z\right) \right] \phi \left( q,z\right)
=q^{2}\phi \left( q,z\right) ,  \label{sleq}
\end{equation}%
which we have cast into Sturm-Liouville form, with the potential 
\begin{equation}
V\left( z\right) =\left( \frac{15}{4}+m_{5}^{2}R^{2}\right) \frac{1}{z^{2}}%
+\lambda ^{2}\left( \lambda ^{2}z^{2}+2\right) .  \label{vsw}
\end{equation}%
For constant $m_{5}$ the solutions of this eigenvalue problem can be found
analytically \cite{for10}. Using Eq. (\ref{m}), the resulting discrete
square-masses $M_{n}^{2}=q_{n}^{2}$ of the $n$-th radial meson excitation
follow the linear trajectories 
\begin{equation}
M_{n}^{2}=4\left( n+\frac{\Delta }{2}\right) \lambda ^{2}  \label{specs}
\end{equation}%
with universal slope $4\lambda ^{2}$, which indeed grow with $\Delta $ \cite%
{for10}. While the (up to a factor) unique quark-antiquark interpolator $J_{%
\bar{q}q}^{A}=\bar{q}^{a}t^{A}q^{a}$ with $\Delta _{\bar{q}q}=3$ corresponds
to ordinary scalar\ mesons, there exist several four-quark interpolators
with $\Delta _{\bar{q}^{2}q^{2}}=6$ \cite{mai07}. Those do not have an 
\textit{a priori} unique assignment to specific meson states with a dominant
tetraquark component. In our case, a suggestive choice may be $J_{\bar{q}%
^{2}q^{2}}^{A}=\varepsilon ^{abc}\varepsilon ^{ade}\bar{q}^{b}C\Gamma ^{A}%
\bar{q}^{c}q^{d}C\Gamma ^{A}q^{e}$ which contains a good diquark and 
antidiquark. For our purposes it suffices, however, to specify the
tetraquark interpolator's quantum numbers, its scaling dimension and the
defining property of maximal overlap with the tetraquark ground state.

As anticipated, Eq. (\ref{specs}) with $\Delta _{\bar{q}^{2}q^{2}}=6$ (i.e.
no anomalous dimension) generates a substantially larger square mass $%
M_{\Delta =6,0}^{2}=2M_{q\bar{q},0}^{2}$ than the $\Delta _{\bar{q}q}=3$
ground state. Hence it misses the exceptionally strong binding required for
a light tetraquark state. In view of our above discussion of the holographic
multiquark representation this is not surprising. Indeed, if the exceptional
lightness of the tetraquark ground state originates from multiquark
correlations (e.g. in the good diquark channel), it should be encoded in the
anomalous dimension $\gamma $ of the tetraquark interpolator which is
neglected in Eq. (\ref{specs}).

Inclusion of this anomalous dimension generalizes Eq. (\ref{m}) to the bulk
mass term 
\begin{equation}
m_{5}^{2}\left( z\right) R^{2}=\left[ 6+\gamma \left( z\right) \right] \left[
2+\gamma \left( z\right) \right]  \label{m2}
\end{equation}%
for the modes dual to the $\Delta _{\bar{q}^{2}q^{2}}=6$ interpolator. This
adds the correction 
\begin{equation}
\Delta V\left( z\right) =\gamma \left( z\right) \left[ \gamma \left(
z\right) +8\right] \frac{1}{z^{2}}  \label{DelV}
\end{equation}%
to the potential (\ref{vsw}) with $m_{5}^{2}R^{2}=12$. Due to the absence of
QCD information on $\gamma $, a physically reasonable ansatz $\gamma 
\left( z\right) =-az^{\eta }+bz^{\kappa }$ is then adopted\footnote{%
Anomalous dimensions of a qualitatively similar $z$ dependence (in the
region of interest) emerge in dual\ backgrounds of holographic RG-flow type 
\cite{mue10}.} and tightly constrained by consistency, stability and physics
requirements. This strategy provides a range of qualitative as well as
quantitative insights \cite{for10} which we summarize in the following.

To begin with, the correction (\ref{DelV}) is bounded by $\Delta V\left(
z\right) \geq -16/z^{2}$ (for any $\gamma $) which prevents the collapse of
the eigensolutions into the AdS$_{5}$ boundary. The bound is saturated by $%
\gamma \equiv -4$ and yields the lower bound%
\begin{equation}
M_{\bar{q}^{2}q^{2},0}\geq M_{\Delta =2,0}=2\lambda  \label{mbd}
\end{equation}%
on the lightest tetraquark mass which Eq. (\ref{m2}) can generate. Moreover,
for constant $\gamma $ in the range $-4<\gamma <-3$ one has $M_{\bar{q}%
^{2}q^{2},0}<M_{\bar{q}q,0}.$ Tetraquark excitations with masses $M_{\bar{q}%
^{2}q^{2},n}$ at or beyond the $M_{\bar{q}q,n}$ (for $n>0$) require a
suitably running $\gamma \left( z\right) $, however, as introduced with the
above ansatz. The latter encodes an exceptionally large binding energy which
drives $M_{\bar{q}^{2}q^{2},0}$ from $\sim 40\%$ above down to $\sim 20\%$
below\footnote{%
Hence the phenomenological ratio $m_{\bar{q}^{2}q^{2},0}/m_{\bar{q}q,0}\sim
0.8/1.5$ is not quantitatively reproduced, perhaps due to the neglect of the
anomalous dimension of the $\bar{q}q$ interpolator or couplings to
additional bulk fields.} the $\bar{q}q$ ground-state mass $M_{q\bar{q},0}=%
\sqrt{6}\lambda $. Around $n\gtrsim 2$, on the other hand, the tetraquark
masses $M_{\bar{q}^{2}q^{2},n}$ start to exceed the corresponding $M_{\bar{q}%
q,n}$. Hence these excitations should become broad enough to prevent
supernumeral scalar states. The above binding mechanism will work similarly
in other AdS/QCD duals, furthermore, since $\gamma $ enters the bulk
dynamics exclusively through the mass term (\ref{m2}) which\ is
model-independently fixed by the AdS/CFT dictionary and generates the
universal correction (\ref{DelV}) in any AdS/QCD dual.

\section{Summary and conclusions}

We have analyzed the holographic representation of multiquark correlations
in hadrons. Our approach is rooted in the observation that the multiquark
content\ of a given hadron reflects itself in the quark composition of the
QCD\ interpolator with which it has maximal overlap. Information on this
substructure is encoded in a gauge-invariant and therefore holographically
active way in the interpolator's anomalous dimension which the AdS/CFT
dictionary translates into a multiquark-content-dependent mass\ correction
for the dual bulk modes. The typically gauge-dependent multiquark
correlations thereby leave a gauge-invariant imprint on the holographic
description.

The implementation of this robust and generic mechanism into two AdS/QCD
duals has yielded the first holographic evidence for important multiquark
correlations in hadrons. To start with, it reveals a strikingly systematic
role of the good-diquark fraction $\kappa $ in baryon spectroscopy and
predicts that the size of light-quark baryons decreases with growing $\kappa 
$. Moreover, it reproduces the masses of all 48 observed nucleon and $\Delta 
$ resonances with just one scale parameter and better accuracy than e.g. any
quark model. In the light scalar meson sector, furthermore, we find
holographic evidence for an exceptionally large four-quark binding energy
and thus for the lightest scalar mesons to consist (mostly) of tetraquarks.
Higher tetraquark excitations can become heavy enough, on the other hand, to
prevent more low-lying scalar resonances than experimentally seen.

\begin{theacknowledgments}
It is a pleasure to thank the organizers for a very informative 
conference and Eberhard Klempt for the 
enjoyable collaboration on diquark effects in the 
baryon spectrum. Financial support from the Deutsche 
Forschungsgemeinschaft (DFG) is also acknowledged.
\end{theacknowledgments}


\end{document}